\documentclass[twocolumn,preprintnumbers,amsmath,amssymb,prl,floatfix]{revtex4}

\usepackage{graphicx}
\usepackage{dcolumn}
\usepackage{bm}

\newcommand{\keV}{{\rm keV}}
\newcommand{\MeV}{{\rm MeV}}
\newcommand{\GeV}{{\rm GeV}}

\newcommand{\Mpc}{{\rm Mpc}}
\newcommand{\kpc}{{\rm kpc}}

\newcommand{\km}{{\rm km}}

\newcommand{\s}{{\rm s}}

\begin{document} 

\title{
$\bm{\gamma}$-ray polarization constraints on Planck scale violations of special relativity}
\author{Luca Maccione, Stefano Liberati, Annalisa Celotti}
\affiliation{SISSA/ISAS, via Beirut 2-4, 34014, Trieste} 
\affiliation{INFN, Sezione di Trieste, via Valerio, 2, 34127 Trieste, Italy}
\author{John G. Kirk}
\affiliation{Max-Planck-Institut f\"ur Kernphysik, Saupfercheckweg, 1, 
D-69117, Heidelberg, Germany}
\author{Pietro Ubertini}
\affiliation{IASF-INAF, via Fosso del Cavaliere 100, Roma, Italy} 
\date{\today} 

\begin{abstract}
Using 
recent polarimetric observations of the Crab Nebula in the
hard X-ray band by INTEGRAL, we show that the absence of vacuum
birefringence effects constrains $O(E/M)$ Lorentz violation in QED to
the level $|\xi| < 9\times10^{-10}$ at $3\sigma$ CL, tightening by
more than three orders of magnitude previous constraints.  We show
that planned X-ray polarimeters have the potential to 
probe $\left|\xi\right|\sim 10^{-16}$ 
by detecting polarization in active galaxies
at red-shift $\sim1$.
\end{abstract}

\maketitle

Experimental constraints on the parameters
quantifying Lorentz invariance violation (LV) are of fundamental
importance. Because the lowest order corrections predicted in the
photon dispersion relation imply the vacuum is birefringent,
observations of polarized photons from distant astronomical sources
provide very promising tests. In this Letter we exploit the
recently discovered linear polarization of hard X-rays from the Crab
Nebula (CN) \cite{integralpol}. These observations show a remarkably high
degree of linear polarization ($46\pm10\%$) and very close alignment
of the polarization vector with both the optical polarization vector
and the projection on the sky of the spin axis of the central neutron
star. The high degree of polarization together with the lack of
detectable rotation of the polarization vector of these $\sim200\,$keV
photons whilst propagating over the intervening $\sim6\times10^{21}$cm
enables us to tighten existing constraints by three orders of
magnitude.

Recent years have witnessed a growing interest 
in 
the possible
high energy violations of local Lorentz Invariance 
as well as a
flourishing of observational tests. Indeed, specific hints of LV arose
from various approaches to Quantum Gravity
\cite{Kostelecky:1988zi,AmelinoCamelia:1997gz,Gambini:1998it,Carroll:2001ws,
Lukierski:1993wx,
AmelinoCamelia:1999pm,Burgess:2002tb,Barcelo:2005fc}.
However, most 
tests 
require a well established theoretical
framework to calculate reaction rates and describe the particle
dynamics.  
Here, we work 
within the framework of Effective
Field Theory with non-renormalizable, mass dimension 5 LV operators
(see~\cite{Mattingly:2005re,Myers:2003fd} and references therein)
restricted to QED, for which the most general dispersion relations for
photons and electrons are
\begin{eqnarray}
\omega_{\pm}^2 &=& k^2 \pm \xi k^3/M
\label{eq:disp_rel_phot}\\
 E_\pm^2 &=& p^2 + m^2 + \eta_\pm p^3/M\;,
\label{eq:disp_rel_ferm}
\end{eqnarray}
where (\ref{eq:disp_rel_phot}) refers to photons~\footnote{This kind
of dispersion relation was also derived in some semi-classical limit
of Loop quantum gravity~\cite{Gambini:1998it}.}  and
(\ref{eq:disp_rel_ferm}) to fermions \footnote{For positrons we have $\eta_{\pm}^{\rm pos} = - \eta_{\mp}^{\rm el}$ \cite{Jacobson:2005bg}.}. 
%
We
assume $M$ to be comparable
to the Planck mass $M_{\rm Pl} \simeq 1.22\times 10^{19}~\GeV$.  The
constants $\xi$ and $\eta_\pm   
$ indicate the
strength of the LV.  The $\pm$ signs denote right and left circular
polarization in (\ref{eq:disp_rel_phot}), and positive and negative
helicity states of the fermion in (\ref{eq:disp_rel_ferm}).
Equation (\ref{eq:disp_rel_phot}) implies that the 
direction of polarization rotates 
during propagation
due to
the different 
velocities of the right- and left-handed
circular polarizations, $v_{\pm} \simeq 1 \pm \xi\, k/M$. This effect
is 
known as vacuum birefringence (VB).

Although it may seem hopeless to search directly for effects
suppressed by the Planck energy scale, even tiny corrections can be
magnified to measurable ones when dealing with high energies, long
distances of signal propagation or peculiar reactions (see,
e.g.,~\cite{Mattingly:2005re, AmelinoCamelia:2002dx}).
Recently $\eta_{\pm}$ have been constrained to have a magnitude less
than 
$10^{-5}$ at 95\% confidence level (CL) by a detailed analysis of
the synchrotron component of the CN broadband spectrum
\cite{Maccione:2007yc}, while the 
constraint $|\xi| \lesssim
2\times 10^{-7}$ has been obtained by \cite{Fan:2007zb} considering
the absence of VB effects during the propagation of
optical/UV polarized light from Gamma-Ray Bursts (GRB)~\footnote{Even
stronger constraints, $O(10^{-14})$, were claimed
in~\cite{Mitro,Jacobson:2003bn} from GRB~021206
observations~\cite{CB}; however the result was later
contested~\cite{RF}.}.  There are also preliminary indications, based
on an analysis of the photon fraction in Ultra-High-Energy Cosmic
Rays, that these coefficients might be less than 
$10^{-14}$, though
nothing conclusive can be claimed yet \cite{Galaverni:2007tq,
Maccione:2008iw}.

In this work we tighten the current constraints on $O(E/M_{\rm Pl})$
suppressed LV by about three orders of magnitude for photons, by
considering the limits on VB effects 
implied by 
the recently detected
\cite{integralpol} polarized hard X-rays from the CN. Firstly, we set
such constraints following the arguments by \cite{Gleiser:2001rm,
  Jacobson:2003bn}, an approach robust against systematic
uncertainties related to astrophysical modeling. We 
then infer 
tighter
limits 
that exploit and rely on modeling of the Crab Nebula
and pulsar.

Finally, we consider the 
constraints which future X-ray
polarization measurements of extragalactic objects, e.g.~Active
Galactic Nuclei (AGN) will allow.  This is of particular
interest 
in the light of current experimental efforts 
to build X-ray polarimeters \cite{Costa:2006cx, Costa:2006pj,
Muleri:2006pg, pogolite}.

During propagation over a distance 
$d$~\footnote{For an
extragalactic object at redshift $z$, the (cosmological) distance is
given by $ d(z) = \frac{1}{H_{0}}\int^{z}_0
\frac{1+z'}{\sqrt{\Omega_{\Lambda} + \Omega_{m}(1+z')^{3}}}\,dz'\;$,
which includes a $(1+z')^{2}$ factor in the integrand to take into
account the red-shift acting on the photon energies. As usual, $H_{0}$
is the present value of the Hubble parameter, while $\Omega_{\Lambda}$
and $\Omega_{m}$ represent the density fractions of cosmological
constant and matter in the Universe, respectively.}, 
the polarization vector of a linearly polarized plane wave with momentum
$k$ rotates through an angle
\cite{Gambini:1998it, Gleiser:2001rm, Jacobson:2003bn,
Jacobson:2005bg},
\begin{equation} 
\theta(k,d) = \frac{\omega_{+}(k)-\omega_{-}(k)}{2}d \simeq \xi\frac{k^2 d}{2\,M_{\rm Pl}}\;.
\label{eq:theta}
\end{equation} 
Observations of polarized light from a distant source can 
constrain $|\xi|$ in 
two ways, depending on the amount of
available information on both the observational and the theoretical
(i.e.~source modeling) side:
\begin{enumerate}
\item
Since detectors have a
finite energy bandwidth, eq.~(\ref{eq:theta}) is never probed in real
situations. Rather, if some net amount of polarization is measured in
the band $k_{1} < E < k_{2}$, an order-of-magnitude constraint arises
from the fact that if the angle of polarization rotation
(\ref{eq:theta}) were to differ by more than $\pi/2$ over this band,
the detected polarization would fluctuate sufficiently for the net
signal polarization to be suppressed \cite{Gleiser:2001rm,
Jacobson:2003bn}.  
From (\ref{eq:theta}), this constraint is
\begin{equation} 
\xi\lesssim\frac{\pi\,M_{\rm Pl}}{(k_2^2-k_1^2)d(z)}
\;,
\label{eq:decrease_pol}
\end{equation} 
%
This just requires that any intrinsic polarization (at source) is not
completely washed out during signal propagation. It thus relies on the
mere detection of a polarized signal, without considering the observed
polarization degree.
A more refined limit 
can be obtained by calculating the maximum
observable polarization degree, 
given the maximum 
intrinsic value \cite{McMaster}:
%
\begin{equation} 
\Pi(\xi) = \Pi(0) \sqrt{\langle\cos(2\theta)\rangle_{\mathcal{P}}^{2}
+\langle\sin(2\theta)\rangle_{\mathcal{P}}^{2}},
\label{eq:pol}
\end{equation} 
where $\Pi(0)$ is the maximum intrinsic degree of polarization,
$\theta$ is defined in eq.~(\ref{eq:theta}) and the average is
weighted over the source spectrum and instrumental efficiency,
represented by the normalized weight function
$\mathcal{P}(k)$~\cite{Gleiser:2001rm}.  
Conservatively, one can set $\Pi(0)=100\%$, but a lower value 
can sometimes be justified on the basis of source modeling.
Using \eqref{eq:pol}, one can then 
cast a constraint by 
requiring $\Pi(\xi)$ to exceed the observed value. 

\item
Suppose 
that
polarized light 
measured in a certain energy band 
has
a position angle $\theta_{\rm obs}$ with respect to a fixed
direction. At fixed energy, the polarization vector rotates by the
angle (\ref{eq:theta}) \footnote{Faraday rotation is negligible at
such energies.}; if the position angle is measured by averaging over a
certain energy range, the final net rotation 
$\left<\Delta\theta\right>$
is given by the
superposition of the polarization vectors of all the photons in that
range:
%
%
\begin{equation}
\tan (2\left\langle\Delta\theta\right\rangle) = \frac{
\left\langle\sin(2\theta)\right\rangle_{\mathcal{P}}}{\left\langle
\cos(2\theta)\right\rangle_{\mathcal{P}}}\;,
\label{eq:caseB}
\end{equation}
where 
$\theta$ is given by (\ref{eq:theta}).
If the position angle at emission 
$\theta_{\rm i}$ in the same energy band 
is known from a model of the emitting source, a constraint can be set by
imposing
\begin{equation}
\tan(2\left\langle\Delta\theta\right\rangle) < \tan(2\theta_{\rm obs}-2\theta_{\rm i})\;.
\label{eq:constraint-caseB}
\end{equation}
%
Although this limit is tighter than that obtained from the
previous methods, it clearly hinges on assumptions about the 
nature of the source, 
which may introduce significant uncertainties.

\end{enumerate}

\begin{figure}[tbp]
\includegraphics[scale=0.4]{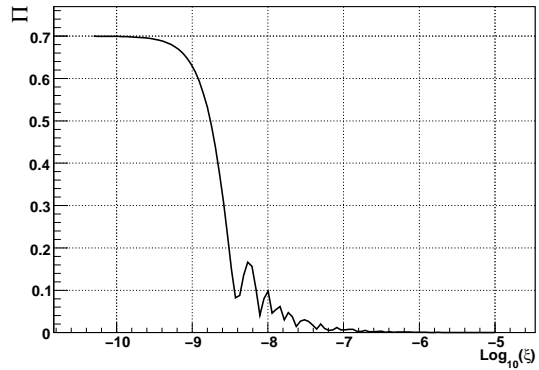} 
\caption{Constraint for the polarization degree. Dependence of $\Pi$
on $\xi$ for the distance of the CN and photons in the 150--300 keV
range, for a constant $\mathcal{P}(k)$.}
\label{fig:caseA}
\end{figure}

In the case of the Crab Nebula, a $(46\pm10)$\%
degree of linear polarization in the $100~\keV - 1~\MeV$ band has 
recently been measured 
by the INTEGRAL mission
\cite{integral,integralpol}.  This 
measurement uses all photons within the SPI instrument energy band. However
the convolution of the instrumental sensitivity to polarization with
the detected number counts as a function of energy, $\mathcal{P}(k)$,
is maximized and approximately constant within a narrower energy band
(150 to 300 keV) and falls steeply outside this range \cite{McGlynn:2007pz}.  For this reason we shall,
conservatively, assume that most polarized photons are concentrated in
this band.
Given $d_{\rm Crab}=1.9~\kpc$, $k_2 = 300~\keV$ and $k_1 = 150~\keV$, 
eq.~(\ref{eq:decrease_pol}) leads to the
order-of-magnitude estimate $|\xi| \lesssim 2\times10^{-9}$.
A more accurate limit follows from 
(\ref{eq:pol}).
In the case of the CN there is
a robust understanding that photons in the range of interest are
produced via the synchrotron proces, for which the maximum degree of
intrinsic linear polarization is about $70\%$ (see e.g.~\cite{Petri:2005ys}).
Figure \ref{fig:caseA} illustrates the dependence of $\Pi$ on $\xi$
for the distance of the CN and for $\Pi(0)=70\%$. 
The requirement $\Pi(\xi)>16\%$
(taking account of a $3\sigma$ offset from the best fit value $46\%$) leads to
the constraint (at 99\% CL)
%
\begin{equation}
|\xi| \lesssim 6\times 10^{-9}\;.
\label{eq:constraint-degree}
\end{equation}
It is interesting to notice that X-ray polarization
measurements of the CN already available in 1978
\cite{1978ApJ...220L.117W}, set a constraint $|\xi| \lesssim
5.4\times10^{-6}$, only one order of magnitude less stringent than
that reported in \cite{Fan:2007zb}.

Constraint (\ref{eq:constraint-degree}) can be tightened
by exploiting the current astrophysical understanding of the source.
The CN is a cloud of relativistic particles and fields powered
by a rapidly rotating, strongly magnetized neutron star. Both the {\em
Hubble Space Telescope} and the {\em Chandra} X-ray satellite have
imaged the system, revealing a jet and torus that clearly identify the
neutron star rotation axis \cite{Ng:2003fm}. The projection of this
axis on the sky lies at a position angle of
$124.0^{\circ}\pm0.1^{\circ}$ (measured from North in
anti-clockwise). The neutron star itself emits pulsed radiation at its
rotation frequency of 30 Hz. In the optical band these pulses are
superimposed on a fainter steady component with a linear polarization
degree of ~30\% and direction precisely aligned with that of the
rotation axis \cite{Kanbach:2005kf}.  The direction of polarization
measured by INTEGRAL-SPI in the $\gamma$-rays is $\theta_{\rm obs} =
123^{\circ}\pm11^{\circ}$ ($1\sigma$ error) from the North, thus also
closely aligned with the jet direction and remarkably consistent with
the optical observations.

\begin{figure*}[tbp]
\includegraphics[scale=0.4]{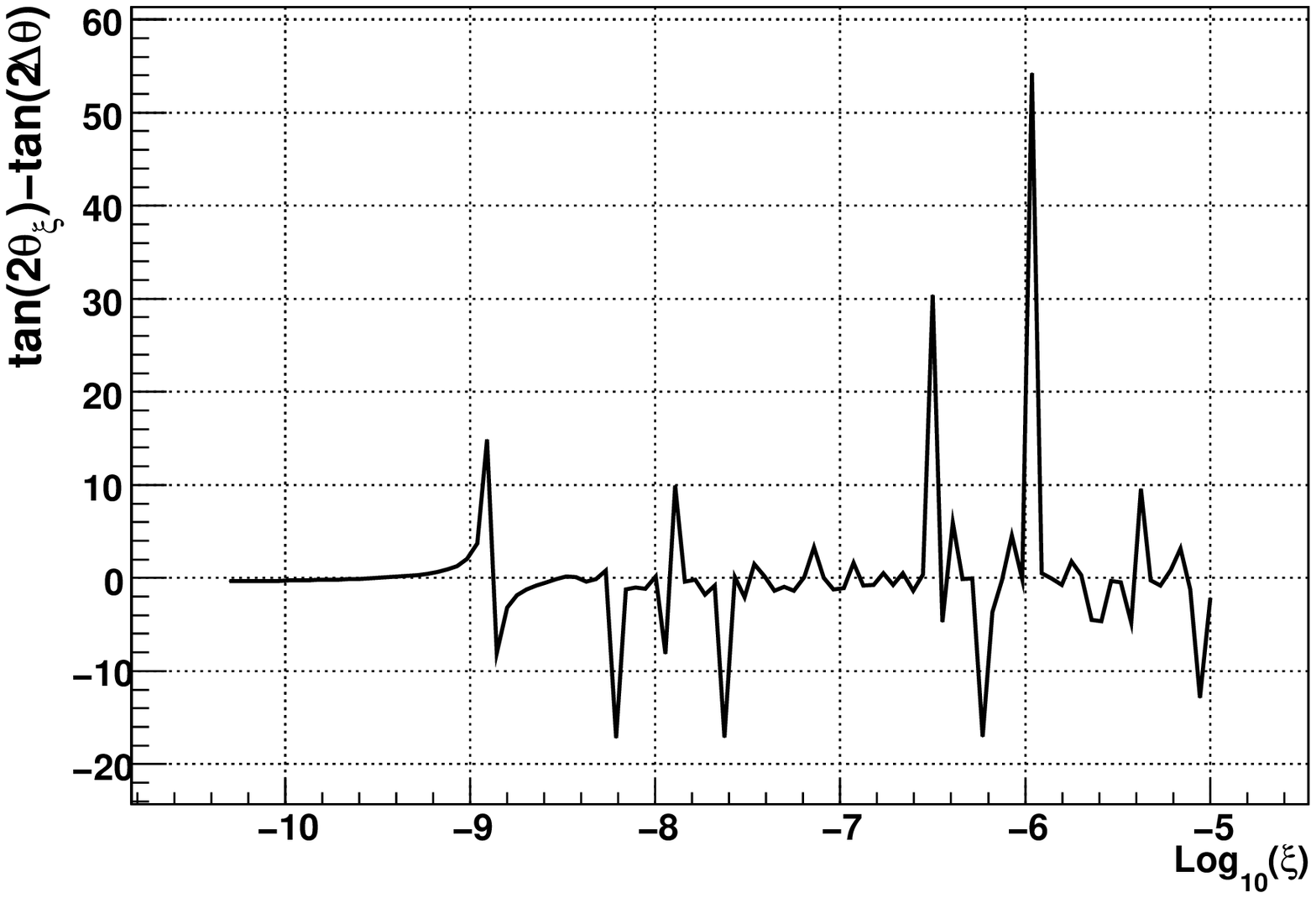} 
\includegraphics[scale=0.4]{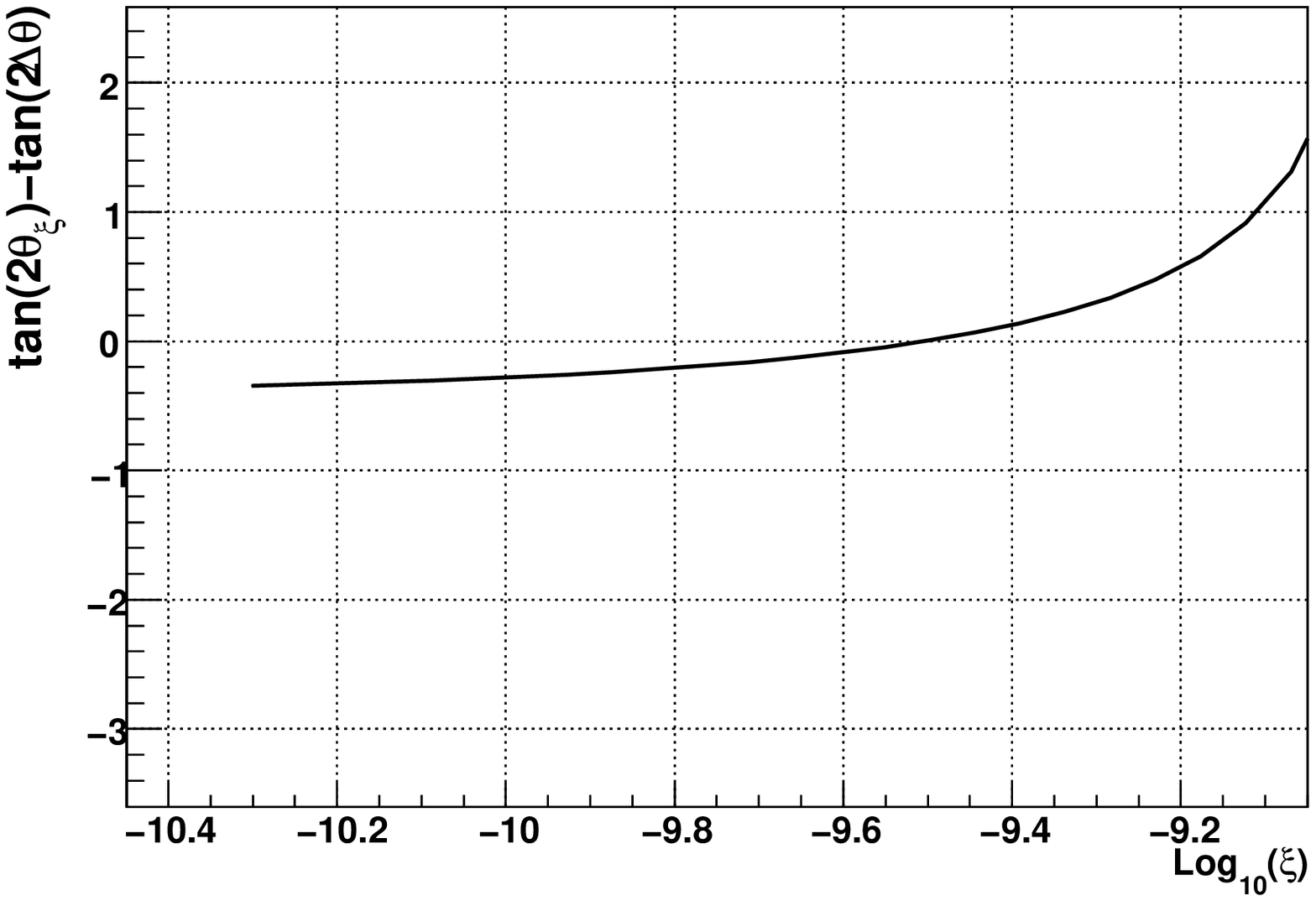} 
\caption{Constraint for the polarization rotation case. Left panel:
dependence of $\tan(2\theta_{\rm f})$ on $\xi$. The spikes correspond
to rotations by $\pi/4$.  Right panel: a zoom-in on the interesting
range of values. The constraint is cast according to
eq.~(\ref{eq:caseB}).}
\label{fig:caseB}
\end{figure*}

This compelling (theoretical and observational) evidence allows us to
use eq.~(\ref{eq:constraint-caseB}). Conservatively assuming
$\theta_{\rm i}-\theta_{\rm obs} = 33^{\circ}$
(i.e.~$3\sigma$ from $\theta_{\rm i}$, 99\% CL), this translates into
the limit
\begin{equation}
|\xi| \lesssim 9\times10^{-10}\;,
\label{eq:constraint-serious-crab}
\end{equation}
and $|\xi| \lesssim 6\times10^{-10}$ for a $2\sigma$ deviation (95\%
CL). 
Figure~\ref{fig:caseB} shows $\tan(2\theta_{\rm f})$
as function of $\xi$. The left--hand panel reports the global
dependence (the spikes correspond to rotations by $\pi/4$), while the
right--hand panel focuses on the interesting range of values \footnote{Note
that the constraint (\ref{eq:constraint-degree}) rules out the
possibility that the polarization angle is close to the expected one
after rotating by some multiple of $\pi$ (the polarization angle is
defined on the interval $[0,\pi]$).}.

\begin{figure*}[tbp]
\includegraphics[scale = 0.4]{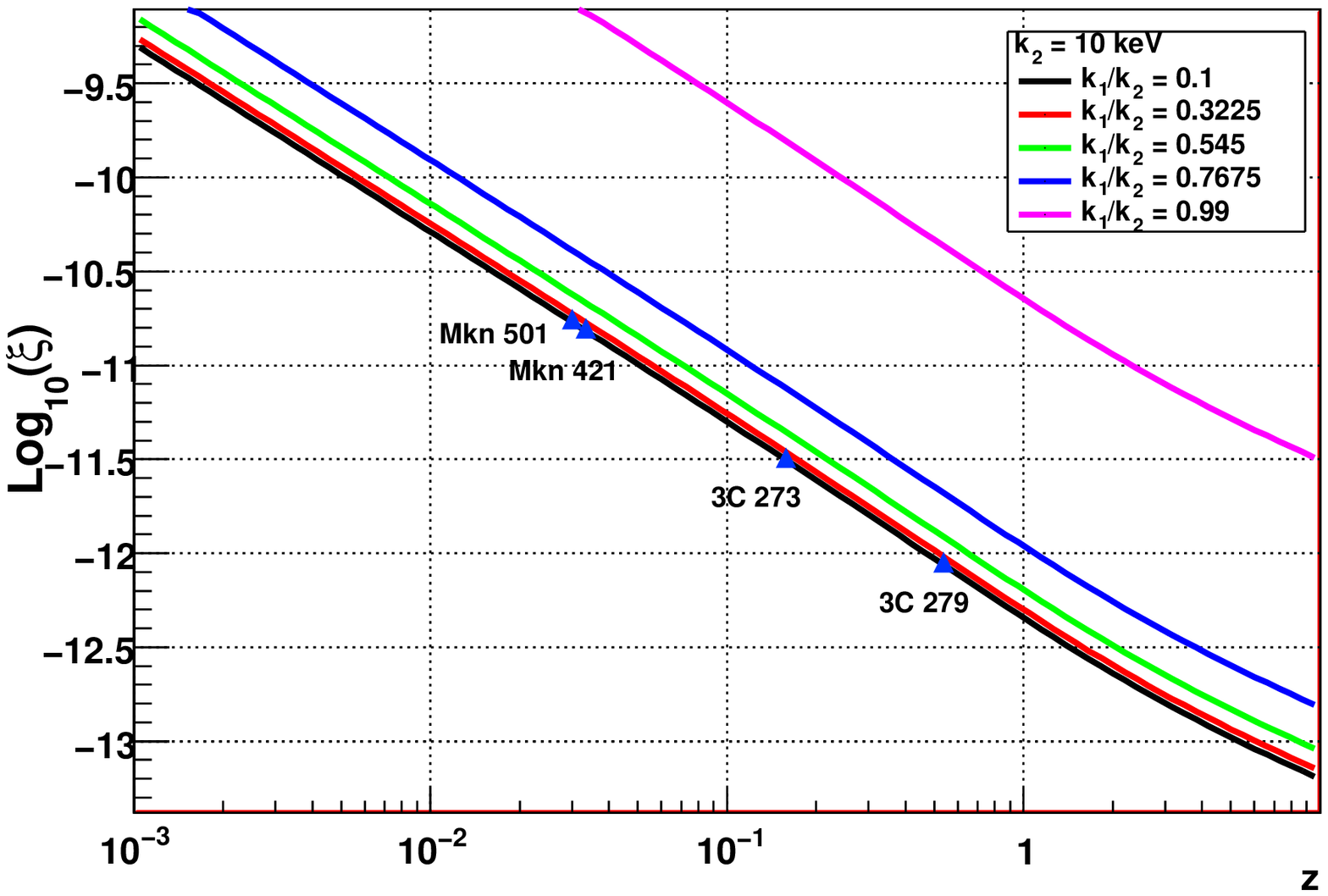}
\includegraphics[scale = 0.4]{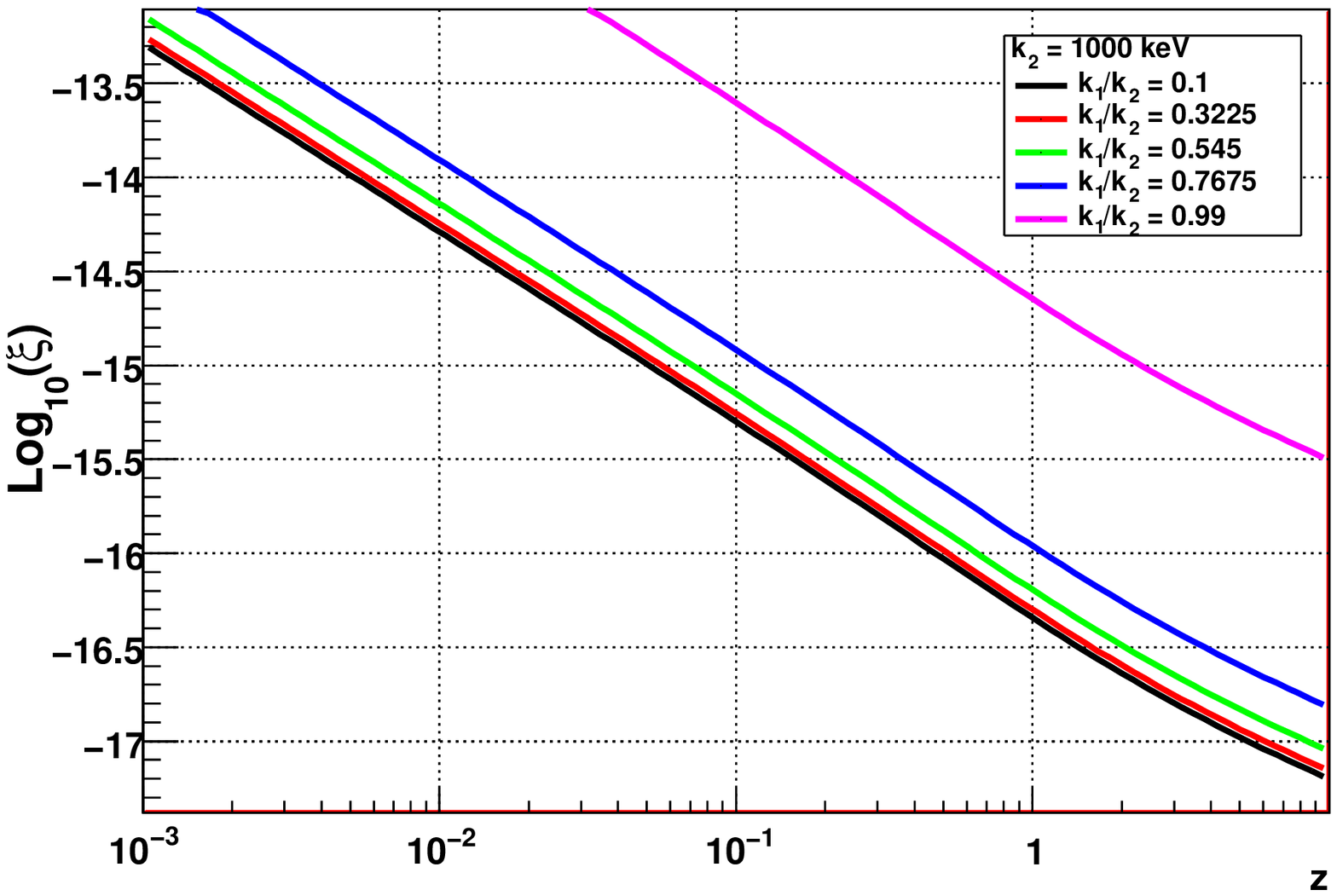}
\caption{Expected constraints from medium X- and soft $\gamma$-ray
polarimetry of extra-galactic sources. High energy scale $k_{2} =
10~\keV$ (left panel) and $1~\MeV$ (right panel), with $\kappa\equiv
k_{1}/k_{2}$ from 0.1 to 0.99. Points in the left panel refer to the
characteristics of a new generation X-ray polarimeter
\cite{Costa:2002yq} assuming that polarization is detected
from the mentioned objects. The constraints are derived as in case of
eq.~(\ref{eq:decrease_pol}) for a concordance cosmology ($\Omega_{m} =
0.28$, $\Omega_{\Lambda} = 0.72$ and $H_{0} = 73~
\km\,\s^{-1}\Mpc^{-1}$).}
\label{fig:plot}
\end{figure*}

The constraints presented in (\ref{eq:constraint-degree})
and (\ref{eq:constraint-serious-crab})
are remarkably strong. Although based on a cumulative effect,
they are achieved using a local (Galactic) object. 
The reason lies, on the one hand, in the quadratic dependence of $\theta$
on the photon energy, in constrast with the linear gain given by
distance (see e.g.~eq.~(\ref{eq:theta})). On the other hand, 
the robust theoretical understanding of the CN has enabled us to 
strengthen the constraints significantly.

Further improvements on 
LV constraints via birefringenge are expected
thanks to the forthcoming high-energy polarimeters, such as XEUS
\cite{Costa:2002yq}, PoGoLite \cite{pogolite}, Polar-X \cite{Costa:2006pj} and Gamma Ray Imager \cite{Knodlseder:2007yj} which will provide an
unprecedented sensitivity, sufficient to detect polarized light at a
few \% levels also in extragalactic sources.
The LV limits will be optimized by balancing between source distance
and observational energy range depending on the detector sensitivity.
This is illustrated in Fig.~\ref{fig:plot}, where the strength of the
possible constraints (cast with the first, most general method
described above) is plotted versus the distance of sources (in
red-shift $z$) and for different energy bands (medium X- and
$\gamma$-rays). Remarkably, constraints of order $|\xi| < O(10^{-16})$
could be placed if some polarized distant sources ($z \sim 1$) will be
observed by such instruments at 1 MeV.


\section*{Acknowledgments}
LM, SL and AC acknowledge the Italian MIUR for financial support. PU
wishes to thank A.J. Dean and J.B. Stephen for useful scientific
discussions.  

\bibliographystyle{apsrev} 

\end{document}